# A comprehensive Fourier ($k$-) space design approach for controllable single and multiple photon localization states

**Subhasish Chakraborty[1], Michael C. Parker[2], Robert J. Mears[3], David G Hasko[1]**

1. Microelectronics Research Centre, Cavendish Laboratory, Department of Physics, University of Cambridge, Madingley Road, Cambridge CB3 0HE, UK

2. Fujitsu Laboratories of Europe Ltd., Columba House, Adastral Park, Ipswich IP5 3RE, UK

3. Pembroke College, Trumpington Street, Cambridge CB2 1RF, UK

A Fourier-space based design approach for the systematic control of single and multiple photon localization states in a 1D lattice is presented. Resultant lattices are aperiodic in nature, such that lattice periodicity is not a useful optimization parameter to achieve novel field localization characteristics. Instead, direct control of field localization comes via control of the Parseval strength competition between the different Fourier components characterizing a lattice. This is achieved via an inverse optimization algorithm, tailoring the aperiodic lattice Fourier components to match that of a target Fourier distribution appropriate for the desired photonic localization properties. We present simulation results indicating the performance of a novel aperiodic lattice exhibiting a doubly-resonant high-Q characteristic.

The control of electromagnetic (EM) wave localization using aperiodic lattices[1] is becoming an increasingly important topic due to potential applications in photonic integrated circuits (PIC's) and dense wavelength division multiplexing (DWDM) systems, where high transmission and high-resolution outputs are desired. In general, when an EM field is localized in a lattice-based cavity, the associated cavity modes exhibit dramatically enhanced transmission in comparison with other frequencies in the vicinity, and hence is the basis for useful filtering functionality. Fundamentally, such localization phenomena of the EM wave can be related to Bragg resonances which are responsible for determining the photon propagation through a lattice. In the simplest case, for a periodic 1D lattice, where the wavelength of the EM radiation is twice the optical lattice constant ($\Lambda$), Bragg-resonance occurs and a bandgap forms in the dispersion characteristic, which results in a dip in the spectral transmission characteristic[2], but no useful photon localization. In order to achieve localized states with a large quality factor, defects (missing or extra scattering sites) must be introduced into the lattice. This process, known as photonic bandgap (PBG) engineering, leads to the appearance of defect (impurity) modes, which in real space determine the localized cavities storing energy at those defect sites[3,4]. In finite lattices, such modes couple to propagating modes to appear as an enhanced transmission at the corresponding resonant frequency[5].

Conventional PBG engineering with single or multiple defects is currently achieved through a combination of either a direct approach of defining a quasi-periodic lattice in real space using constraining formulae, such as derived from Fibonacci, Cantor or Thue-Morse series[6-8]; coupled-cavity structures, proposed by Yariv *et al.*[9]; or by intuition (accumulated design experience) and trial-and-error, such as the high-Q cavity structures reported by Akahane *et al.*[10]. These methods, which render the lattice aperiodic in the most general sense, indicate that aperiodic lattices offer an important platform for the study of EM wave localization. However, the problem of finding the general selection rule for choosing the relatively small number of aperiodic lattices with useful localization properties, from the very large number of possible aperiodic lattices, is computationally very demanding. In this letter, we present a more efficient approach to the identification of those useful aperiodic lattices by primarily considering the Fourier-space (*k*-space) distribution appropriate for the desired EM wave localization property, rather than direct consideration of the real-space lattice.

The localization of the EM wave is determined by those wavevectors of the EM wave spectrum, which have their Bragg resonating partner in the Fourier spectrum of the scattering dielectric function. In the general case (i.e. in addition to the conventional periodic case) a Bragg resonance is the interaction of an EM wavevector $k$ with a Fourier frequency $G_q$ according to $k = G_q/2$. The strength of the Bragg resonance at that frequency is controlled by the Fourier coefficient amplitude $\bar{\varepsilon}\{G_q\}$ given by the Fourier transform (FT) of the scattering dielectric function $\varepsilon\{x_p\}$. Localization of the EM wave inside a lattice can then be thought of as originating from the interaction of different Bragg resonances, i.e., those finite Fourier components characterising the lattice. Multiple Bragg resonances can therefore be used as the basis of controlling EM localization in a fashion not available from a conventional periodic lattice (which exhibits only a single Bragg resonance). The FT-basis of the localization phenomenon, as qualitatively outlined above, allows us the additional benefit of exploiting digital signal processing techniques[11]. For example, scattering sites of a lattice, doped with defects, can be treated as samples with a spatial frequency of less than the Nyquist frequency (i.e. highest spatial frequency of the lattice). The Nyquist frequency is denoted by the symbol $G_B$, determined by the reciprocal of the minimum optical path-length $\Lambda$ (optical path-length is given by $x = \int n(l)dl$ where $n$ represents refractive index distribution and $l$ is geometric real-space) between adjacent scatterers, $G_B=2\pi/\Lambda$, and determines the conventional Bragg frequency $\omega_B=cG_B/2$, where $c$ is the vacuum speed of light. A discrete Fourier Transform (DFT) of such a (purely-real) distribution produces a symmetric Fourier spectrum about $G_B$, where each component is responsible for a Bragg resonance. Due to considerations of power conservation (i.e. Parseval's theorem) the wave localization properties of the lattice are determined by the strength competition between those Fourier components.

For a binary relative permittivity lattice structure $\varepsilon\{x_p\}$ containing $N$ sites, located at the set of positions $\{x_p\}$ in real-space, a periodic boundary condition provides the corresponding set of positions $\{G_q\} = (2q/N)(2\pi/\Lambda) = (2q/N)G_B, \quad q = 1,2,.....N/2$, in Fourier-space. The maximum value of only $q=N/2$ reflects the symmetric redundancy in the spectrum about the Nyquist frequency, discussed above. We note that $G_B \equiv G_{N/2}$. The DFT of the lattice function $\varepsilon\{x_p\}$ yields the set of discrete Fourier components $\bar{\varepsilon}\{G_q\}$ as:

$$\bar{\varepsilon}\{G_q\} = \frac{1}{N}\sum_{p=0}^{N-1}\varepsilon\{x_p\}e^{-iG_q x_p} \ . \tag{1}$$

In this letter, our approach to the design of an aperiodic lattice, with precisely defined wave localization properties, is to manipulate the relative strength of the different Fourier components (i.e. different Bragg resonances) of the lattice spectrum, (Eq. 1), to match that of a target Fourier component distribution $\bar{\varepsilon}_{\mathrm{target}}\{G_q\}$, appropriate for the desired localization phenomenon. The required real-space lattice configuration is considered 'optimized' (i.e. it has the desired EM wave localization property) when the cost function $E$ is minimized, where

$$E(\varepsilon) = \sum_{q=1}^{N/2}\left[\left|\bar{\varepsilon}_{\mathrm{target}}\{G_q\}\right| - \left|\bar{\varepsilon}\{G_q\}\right|\right]^2 , \tag{2}$$

describes the 'error' between the lattice Fourier spectrum and the target Fourier spectrum $\bar{\varepsilon}_{\mathrm{target}}\{G_q\}$. In order to manipulate the Fourier components distribution, given by equation 1, the real-space lattice periodicity is broken by the introduction of defects in at least one of the sites, producing a trial lattice $\varepsilon_{\mathrm{trial}}\{x_p\}$ with a modified spectral response $\bar{\varepsilon}_{\mathrm{trial}}\{G_q\}$. A new cost function $E(\varepsilon_{\text{trial}})$ is calculated, and the trial lattice $\varepsilon_{\text{trial}}\{x_p\}$ is evaluated in an annealed, probabilistic fashion. This amendment to the lattice is accepted over the previous ($i^{\text{th}}$) configuration lattice $\varepsilon_i\{x_p\}$ if the Boltzmann probability distribution $P(\Delta E)$ exceeds a random number $r$ between 0 and 1, with $P(\Delta E) = \exp(-\Delta E / T)$, $\Delta E = E(\varepsilon_{\text{trial}}) - E(\varepsilon_i)$, and $T$ is the analogous system temperature. Therefore negative values for $\Delta E$ are always accepted. The simulated annealing (SA) algorithm requires the system to start at a high temperature (found by trial-and-error) to ensure that changes causing an increase in $\Delta E$ initially tend to be accepted, so avoiding the problem of becoming trapped in local minima during the early stages of the annealing process. As the algorithm iterates, the temperature is gradually lowered for the $l^{\text{th}}$ iteration according to $T_{l+1} = \alpha T_l$, where $\alpha\,(<1)$ is the cooling rate. As the system cools the probability of accepting positive changes in $\Delta E$ reduces, and the system tends towards a global minimum in cost space. We note that multiple runs of the SA algorithm will tend to find different solutions, each of which is close to an overall global optimum in cost-space. However, from a practical point of view, the functionality of these solutions tends to be indistinguishable.

As a design example, we have chosen to modify the spectral response, reported in Foresi *et al.*[5], from a single resonant high-Q defect state to a doubly resonant system with two high-Q defect states within a wide photonic stop band. In order to achieve this modification in the same system, we need to identify a real-space lattice with two properties; first, the lattice must have same number (eight) of refractive index contrast elements (e.g. etched air holes in a Si-waveguide, i.e. $N$=16); second, the length of any defect is 0.5Λ. We have also used silicon-silicon dioxide as our material system, which provides a sufficiently large contrast (the refractive index difference $\Delta n$ is about 2) so that the optical wave is strongly confined. The waveguide cross-section can therefore also be made very small (in this case 0.5 μm wide and 0.26 μm thick), which is useful in microphotonic integrated circuits at optical telecommunication wavelengths, which require microscale optical elements. The $k$-space design was carried out using a software program written in MATLAB™, taking advantage of the efficient FFT algorithm, the SA optimisation took less than 1 second using a Pentium-IV processor with 2.8GHz clock frequency, and 512MB RAM. The resulting device transmissivities in Fig. 1(c) was simulated using a commercial software package (FIMMPROP-3D), discussed in detail elsewhere[12].

Fig. 1 (i-a) shows the target DFT spectrum appropriate for a single high-Q resonant state within the photonic stop band, as seen in Fig. 1 (i-c). The Fourier amplitude is zero at the Nyquist frequency point $G_B$, with the strongest Fourier components existing at the neighbouring Fourier positions, calculated as $0.875G_B$ with its symmetric partner at $1.125G_B$. The Bragg resonances, which originate from these main two Fourier components, interfere to form a very narrow (i.e. high-Q) transmission peak at the conventional Bragg frequency $\omega_B$. Fig. 1 (ii) & (iii) show the DFT responses and the spectral transmission characteristics of two further lattices doped with multiple defects. The defects control the strength of the Fourier component at the Nyquist frequency, and hence the transmissivity at the Bragg-frequency. In accordance with Parseval's theorem, we note that the Fourier component at $G_B$ in Fig.1 (ii-a) & (iii-a) strengthens at the cost of the component amplitudes of the neighbouring Fourier locations. The strengths of the Fourier components (i.e. closely equivalent to reflection coefficients) determine the finesses associated with the two resonant peaks, and combined with cavity losses (e.g. lattice absorption, diffractive radiation) control the overall Q-values of the peaks. By inspection, the individual resonances in Fig.1 (ii-c) & (iii-c) have lower finesses (and hence Q-values) than the single resonant peak of Fig.1 (i-c). Using our understanding of the Parseval competition process for systematic control of the strengths of the Fourier components, we now introduce a target DFT spectrum $\bar{\varepsilon}_{\text{target}}\{G_q\}$, shown in Fig. 1 (iv-a), appropriate to form two high-Q transmission peaks. Fig. 1 (iv-b) shows the resulting optimal aperiodic lattice solution for this problem. With two defects of size $d$=0.5Λ, the two resonating defect states appear deep within the stop band at frequencies $0.92\omega_B$ and $1.05\omega_B$ (the asymmetry apparent with respect to the Bragg frequency being due to the additional lattice dispersion as calculated by FIMMPROP3D). Evident are the higher finesses of the peaks of Fig.1 (iv-c) as compared with Fig.1 (ii-c) & (iii-c). We also observe an interesting inverse relationship between the relative distance separating the spatial defects, and the frequency distance between the resulting resonating peaks. Such behaviour is perhaps to be expected due to the reciprocal symmetries between the FT conjugate planes. The simulated field intensity patterns in the *xy*-plane corresponding to these modes and at the Bragg frequency have been plotted in Fig. 2. At the resonating frequencies each defect site acts as a cavity, where the electromagnetic wave is spatially localized. However, the evanescent tail of each mode can overlap and couple to the propagating modes resulting in an enhanced transmission for the corresponding resonant frequency. It also can be seen from Fig 2., that although the spatial field distributions for both frequencies show maxima at the defect regions, there is a node between the defects for the low frequency resonance (Fig 2 (a)). As might be expected, the symmetries exhibited by the spatial field distributions at the resonating frequencies have a close analogy with the bonding and antibonding wavefunctions well known in solid-state physics. At the Bragg frequency the high intensity of light entering from the left, but being reflected (i.e., not allowed to propagate) is clearly evident.

In conclusion, we have presented a comprehensive Fourier-space based design approach for systematic control of the number of resonant states within a photonic stop band. We have demonstrated that by manipulating the Fourier components of the lattice under the Parseval constraint, we are able to achieve a desired localization charateristic. As an example, we have presented a design of a doubly resonant microcavity structure using a high refractive contrast material system. The optimum structure is a short and overall aperiodic lattice not known before. Such structures may find a range of useful spectroscopic and PIC applications, e.g. ultrafast optical switching and modulation.

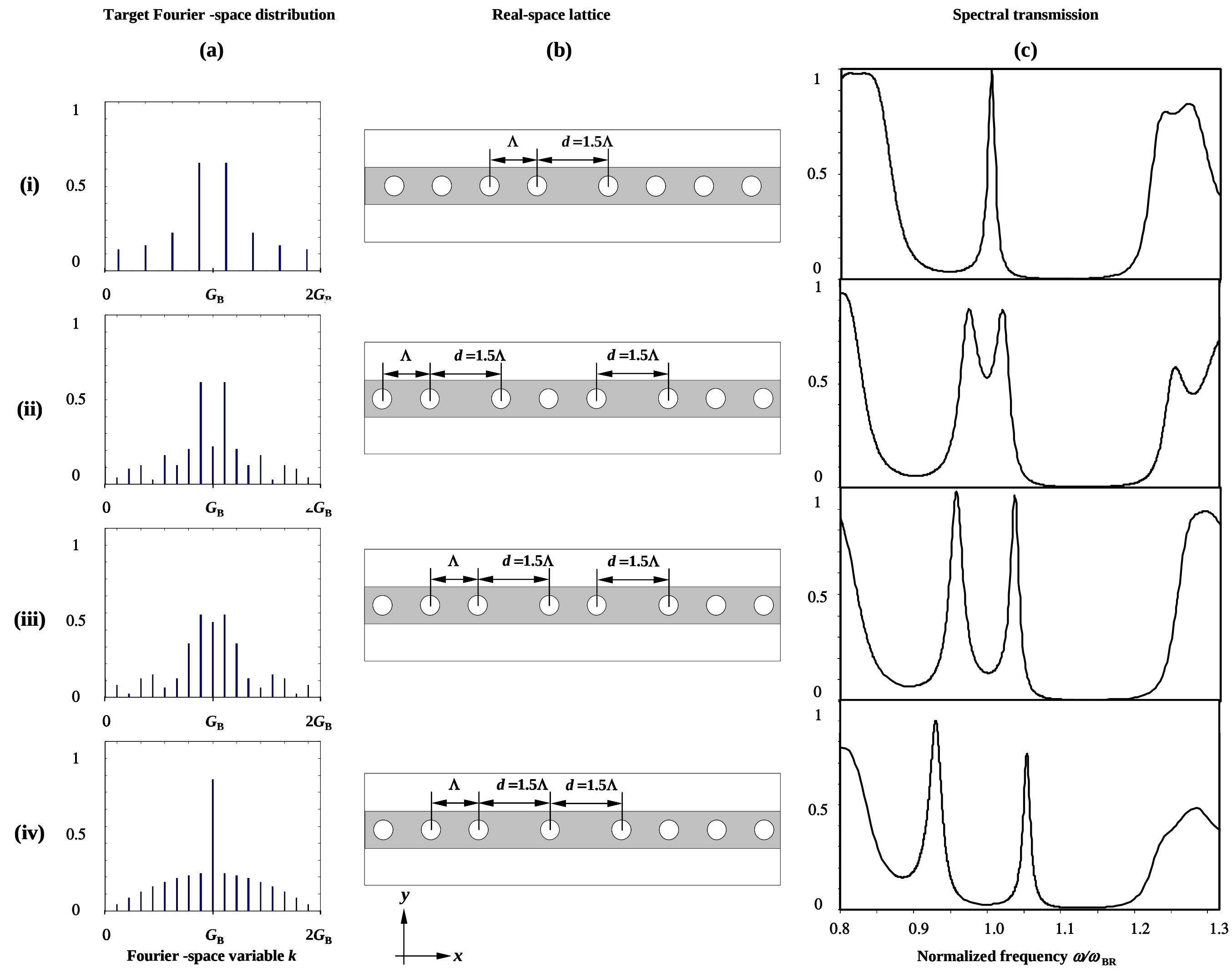

Fig.1.The target Fourier components distributions and spectral transmission characteristics of four different lattices doped with defects. Evident is the systematic progression from a single high-Q transmission peak to double high-Q transmission peaks.

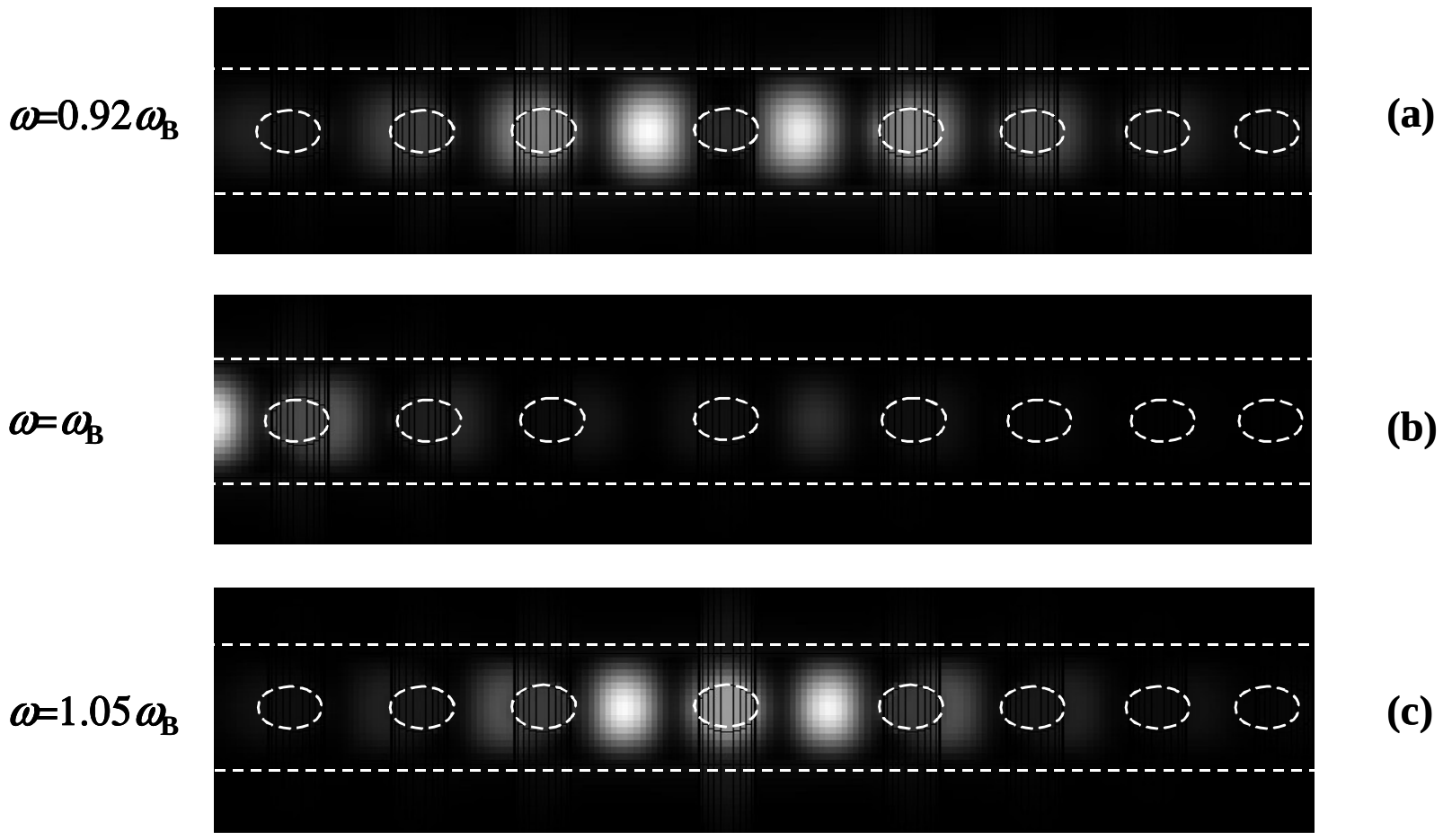

Fig.2. Simulated field intensity patterns in the *xy*-plane corresponding to the resonating modes (at frequencies $\omega$= 0.92$\omega_B$ and 1.05$\omega_B$) and at the Bragg frequency $\omega_B$ of Fig.1 (iv-c).